\documentclass[a4paper,11pt]{article}

\usepackage{jheppub}

\usepackage[utf8]{inputenc}
\usepackage{graphicx}

\usepackage{amssymb}
\usepackage{amsxtra}
\usepackage{amsmath}
\usepackage{booktabs,multirow,tabularx}
\usepackage{slashed}
\usepackage{float}
\usepackage{placeins}
\usepackage{rotating}
\usepackage{lscape}
\usepackage{color}
\usepackage{hyperref}
\usepackage{cancel}
\usepackage[Q=yes,pverb-linebreak=no]{examplep}
\usepackage{ulem}

\DeclareGraphicsExtensions{.pdf}
\graphicspath{{./figures/}}

\newcommand{\lsim}{\lesssim}

\newcommand{\ord}[1]{\mathcal{O}{(#1)}}
\newcommand{\eq}[1]{Eq.~(\ref{#1})}
\newcommand{\fig}[1]{Fig.~(\ref{#1})}

\newcommand{\gev}{\,\textrm{GeV}}

\newcommand{\tev}{\,\textrm{TeV}}

\def\beq{\begin{equation}}
\def\bea{\begin{eqnarray}}
\def\eeq{\end{equation}}
\def\eea{\end{eqnarray}}
\def\beqal{\begin{align}}
\def\endal{\end{align}}

\definecolor{applegreen}{rgb}{0.55, 0.71, 0.0}
\definecolor{purple}{rgb}{0.5,0.0,0.5}

\newcommand{\rmtodo}[1]{{\color{blue} #1}}

\makeatletter
\newcommand\footnoteref[1]{\protected@xdef\@thefnmark{\ref{#1}}\@footnotemark}
\makeatother

\DeclareFontFamily{U}{cbgreek}{}
\DeclareFontShape{U}{cbgreek}{m}{n}{
        <-6>    grmn0500
        <6-7>   grmn0600
        <7-8>   grmn0700
        <8-9>   grmn0800
        <9-10>  grmn0900
        <10-12> grmn1000
        <12-17> grmn1200
        <17->   grmn1728
      }{}
\DeclareFontShape{U}{cbgreek}{bx}{n}{
        <-6>    grxn0500
        <6-7>   grxn0600
        <7-8>   grxn0700
        <8-9>   grxn0800
        <9-10>  grxn0900
        <10-12> grxn1000
        <12-17> grxn1200
        <17->   grxn1728
      }{}

\makeatletter
\newcommand{\normalorbold}{%
  \ifnum\pdf@strcmp{\math@version}{bold}=\z@ bx\else m\fi
}
\makeatother

\begin{document}

\title{Lepton-Flavor-Violating ALPs at the Electron-Ion Collider: A Golden Opportunity}

\author[a]{Hooman Davoudiasl,}

\author[b]{Roman Marcarelli,}

\author[b]{and Ethan T.~Neil}

\affiliation[a]{High Energy Theory Group, Physics Department Brookhaven National Laboratory, Upton, NY 11973, USA}

\affiliation[b]{Department of Physics, University of Colorado, Boulder, Colorado 80309, USA}

\emailAdd{hooman@bnl.gov}
\emailAdd{roman.marcarelli@colorado.edu}
\emailAdd{ethan.neil@colorado.edu}

\abstract{Axion-like particles (ALPs) arise in a variety of theoretical contexts and can, in general, mediate flavor violating interactions and parity non-conservation.  We consider lepton flavor violating ALPs with GeV scale or larger masses which may, for example, arise in composite dark sector models.  We show that a future Electron-Ion Collider (EIC) can uncover or constrain such ALPs via processes of the type $e \, A_Z \to \tau \, A_Z\, a$, where $A_Z$ is a nucleus of charge $Z$ and $a$ is an ALP in the range $m_\tau \leq m_a \lsim 20~{\rm GeV}$.  The production of the ALP can have a large $Z^2$ enhancement from low $Q^2$ electromagnetic scattering of the electron from a heavy ion.  Using the gold nucleus ($Z=79$) as an example, we show that the EIC can explore  
$e-\tau$ flavor violation, mediated by GeV-scale ALPs, well beyond current limits.  Importantly, the EIC reach for this interaction is not sensitive to the lepton-flavor conserving ALP couplings, whose possible smallness can render searches using $\tau$ decays ineffective.   We also discuss how the EIC electron beam polarization can provide a powerful tool for investigating parity violating ALPs.}

\maketitle

\section{Introduction\label{sec:intro}}

Global symmetries arise in a variety of theoretical settings and their spontaneous breaking generally leads to axion-like particles (ALPs) \cite{Peccei:1977hh, Weinberg:1977ma,Wilczek:1977pj,Georgi:1986df}.  Often the global symmetry is not exact and its small explicit breaking leads to the appearance of relatively light ALPs.  In the Standard Model (SM), the spontaneous breaking of chiral symmetries gives rise to pseudo-Nambu-Goldstone bosons, {\it i.e.} the parity-odd pions, which have axion-like properties.  The small masses of light quarks, compared to typical hadronic scale $\Lambda_{\rm QCD}\sim 200$~MeV, provide explicit chiral symmetry breaking, leading to relatively small pion masses. 

One may expect similar phenomena to arise in new sectors of physics, which could provide answers for open questions like the nature of dark matter (DM), for example.  New physics sectors at scales of $\ord{\rm GeV}$ or less have been considered as potential alternatives to beyond SM (BSM) models at or above the weak scale $\sim 100$~GeV.  In particular, ``dark sectors" that could include DM and other  related states and interactions have been extensively studied over the last several years.  In these setups, new experimental possibilities for discovery of BSM  phenomena open up.  Due to the relatively low scale and feeble couplings associated with such ``dark sectors," intense sources can provide good prospects for uncovering them.   

Searches for lepton flavor violation (LFV) mediated by ALPs can be a particularly sensitive probe of new physics \cite{Wilczek:1982rv,Ema:2016ops,Bauer:2019gfk,Cornella:2019uxs,Endo:2020mev,Iguro:2020rby,Han:2020dwo,Mukaida:2021sgv,Bauer:2021mvw}.  For an example of a concrete UV-complete model in which GeV-scale ALPs with lepton flavor violation emerge out of a dark sector giving rise to dark matter and neutrino mass, see Refs.~\cite{Davoudiasl:2017zws}.  The dominant current constraints on flavor off-diagonal ALP-lepton couplings universally come from low-energy experimental searches involving flavor-violating lepton decays \cite{Cornella:2019uxs,Bauer:2021mvw}, such as $\tau \rightarrow e \mu \mu$ for the $a e \tau$ interaction, where $a$ denotes the ALP.  Searches for ALPs produced in Higgs decays can also provide relevant limits \cite{Davoudiasl:2021haa}, particularly for relatively heavy ALP masses above 2 GeV.  However, these experimental probes require the existence of other significant couplings - either flavor-diagonal interactions $a \ell \ell$ or ALP-Higgs couplings - in addition to the flavor-violating couplings.  In particular, the constraints from lepton decays become sharply weaker for axion masses above the tau lepton mass, where the decays become highly suppressed due to kinematics.

  One may also consider $Z$ decays as a probe of ALP mediated LFV at the LHC.  Assuming the $a$-$e$-$\tau$ coupling $C_{\tau e}/\Lambda \lsim 10$ TeV$^{-1}$ (typical of the  parameter space in our analysis below) only and for $m_a \ll m_Z$, we estimate that gives a branching fraction 
\beq
{\rm Br}(Z \rightarrow a e\tau\to e\tau e\tau) \approx {\rm Br}(Z \rightarrow (e^+e^-, \tau^+\tau^-)) \times \frac{C_{\tau e}^2 m_\tau^2}{\Lambda^2} \frac{1}{8\pi^2} \lsim 3\times 10^{-7}\,,
\label{Zto2(etau)}
\eeq
for which no experimental bounds exist at the present 
(the current measurement of 4-lepton $Z$ decays has an uncertainty of $\sim 2\times 10^{-7}$ \cite{ParticleDataGroup:2020ssz}, but only applies to electron and muon final states).  Though not directly relevant, we also note that the limit on flavor-violating $Z \to e \tau$ decay branching fractions is $5 \times 10^{-6}$ at 95\% confidence level \cite{ParticleDataGroup:2020ssz}.  In any event, it could be interesting to consider bounds from $Z$ decays in more detail in future work.

Motivated by the above considerations, in this work, we examine the possibility of probing lepton-flavor-violating ALPs arising at or above the GeV scale, perhaps as part of a new dark sector of physics, in electron-ion collisions at the future Electron-Ion Collider (EIC).  We will focus specifically on probing the $a e \tau$ interaction through the process $e \rightarrow \tau a$.  In contrast with bounds on this interaction from rare lepton decays, the EIC is directly sensitive to the $a e \tau$ coupling with no requirement of significant flavor-diagonal lepton couplings.  Compared to electron-proton scattering experiments, electron-ion collisions offer a coherent enhancement in scattering rate due to the ion charge, increasing the ALP production rate in electromagnetic scattering processes by orders of magnitude (although this enhancement is partially compensated by reduced ion luminosity compared to electron-proton operation.)  These features, along with its high center-of-mass energy, make the EIC uniquely effective at searching for LFV ALPs in theoretically interesting regions of parameter space.  Additionally, the use of beam polarization at the EIC provides an experimental handle on the parity-violating angle in the $a e \tau$ interaction.

In this paper, we will study the ALP EFT allowing for both LFV and parity non-conservation.  Exploiting the presence of LFV, we will focus on tau lepton final states.  Under a wide range of efficiency assumptions, we will estimate projected limits on the $a e \tau$ interaction and compare with existing bounds.  Other works that consider exploring LFV using electron beam facilities include Ref.~\cite{Gonderinger:2010yn,Cirigliano:2021img,Zhang:2022zuz} (EIC) and Ref.~\cite{Furletova:2021wyq} (CEBAF), where a leptoquark sector is generally assumed to mediate the effects studied therein.  Recent work \cite{Liu:2021lan} has also studied the prospects for detecting ALPs through their photon coupling at the EIC.

\section{ALP Effective Field Theory\label{sec:EFT}}
The flavor-violating ALP effective Lagrangian we will adopt for this analysis has been previously considered in Refs. \cite{Bauer:2017ris,Cornella:2019uxs,Bauer:2019gfk,Bauer:2020jbp,Chala:2020wvs,Escribano:2020wua,Calibbi:2020jvd,Ma:2021jkp, Bauer:2021mvw}. It is given by
\begin{align}
    {\cal L} &= \frac{1}{2}(\partial_\mu a)^2 - \frac{1}{2}m_a^2a^2 + {\cal L}_\ell + \cdots + h.c.
\intertext{We will be focusing on the term }
    {\cal L}_\ell &= \frac{\partial_\mu a}{\Lambda}\sum_{\ell\ell'}\bar{\ell}\gamma^\mu \left(V_{\ell\ell'}+A_{\ell\ell'}\gamma_5\right)\ell' + h.c.
\end{align}
where $\Lambda$ is the EFT scale of the ALP theory.  Here, we will take $V_{\ell\ell'}$ and $A_{\ell\ell'}$ to be real, so that the Lagrangian is CP-even. However, note that the presence of the vector-like $V_{\ell\ell'}$ term implies that parity violation is present in general. The parity violation can be parametrized by an angle $\theta_{\ell\ell'}$ by defining $\theta_{\ell\ell'} = -\tan^{-1}(V_{\ell\ell'}/A_{\ell\ell'})$ and $C_{\ell\ell'} = \sqrt{V_{\ell\ell'}^2 + A_{\ell\ell'}^2}$, so that
\bea
{\cal L}_{\ell} = \frac{C_{\ell\ell'}}{\Lambda}\partial_\mu a\sum_{\ell\ell'}\bar{\ell}\gamma^\mu \left(\sin{\theta_{\ell\ell'}}-\cos{\theta_{\ell\ell'}}\gamma_5\right)\ell' + h.c.
\eea
Note that $\theta_{\ell\ell'} = 0$ leaves only the parity-even term, while $\theta_{\ell\ell'} = \pi/4$ ($\theta_{\ell \ell'} = 3\pi/4$) is maximally parity-violating, because then the ALP only interacts with left-handed (right-handed) particles.

It is useful to rewrite the leptonic Lagrangian by integrating by parts and solving the classical equations of motion on the leptons. Doing so, we have
\bea
    {\cal L}_\ell = \frac{C_{\ell\ell'}}{\Lambda}a\sum_{\ell\ell'}\bar{\ell}\left(m^-\sin{\theta_{\ell\ell'}} - m^+\cos{\theta_{\ell\ell'}}\gamma_5\right)\ell' + h.c., \label{eq:PV_lagrangian}
\eea
where $m^\pm\equiv m_\ell\pm m_{\ell'}$.
Here, we note that there is no PV contribution for the flavor-diagonal case, because for $\ell = \ell'$, the difference in masses goes to zero, so we can absorb $\cos \theta_{\ell \ell}$ into the definition of $C_{\ell \ell}$.  Although we will neglect $\theta_{\ell \ell}$, it is worth noting that the presence of parity violation will tend to suppress the diagonal couplings $|C_{\ell \ell}|$ relative to the off-diagonal couplings $|C_{\ell \ell'}|$.  We will retain the explicit dependence on parity violation in the off-diagonal couplings, although we will find that this dependence vanishes in our signal cross-section.

The above model could lead to an interesting LFV (and potentially parity violating) signal at the EIC.  In particular, one can consider the process in \fig{fig:diagram}, where the coupling $C_{\tau e}$ allows for the process $eA_Z \rightarrow \tau A_Z a$, emission of an ALP in which the beam electron is converted to a tau lepton.  The emitted $a$ can then decay into leptonic final states.  

We restrict our attention to the $\tau$ lepton because of the mass-dependence of the ALP coupling: since $m_\tau \gg m_\mu \gg m_e$, the branching fractions into final states containing only $\mu$ and $e$ are negligible.  We further neglect the $C_{\tau \mu}$ coupling, for two reasons: first, the $C_{\tau e}$ coupling is essential for the production of our ALP signal, whereas $C_{\tau \mu}$ will only be relevant for ALP decays; second, current plans for the EIC detectors do not necessarily include muon chambers, in which case muonic decays of the ALP will be difficult to detect. In this scenario, the only terms in the Lagrangian we are interested in are those which contain at least one $\tau$; in particular,
\bea
    {\cal L}_{\tau} &\approx& \frac{C_{\tau e}m_\tau}{\Lambda} a \bar{\tau}(\sin{\theta_{\tau e}} 
    - \cos{\theta_{\tau e}}\gamma_5)e \nonumber \\
    &+& \frac{C_{\tau\tau}m_\tau}{\Lambda} a \bar{\tau}\gamma^5\tau + h.c.\,,
\eea
where $m_e$ is ignored. For $m_a > m_\tau \gg m_e$, the decay rates of interest are
\begin{align}
    \Gamma(a\rightarrow \tau^+\tau^-) &= \frac{|C_{\tau\tau}|^2}{2\pi }\frac{m_\tau^2}{\Lambda^2}\sqrt{m_a^2 - 4m_\tau^2}\\
\intertext{and}
    \Gamma(a\rightarrow \tau^\pm e^\mp) &\approx \frac{|C_{\tau e}|^2}{8\pi}\frac{m_\tau^2}{\Lambda^2}\frac{(m_a^2 - m_\tau^2)^2}{m_a^3}.
\end{align}
For the region of parameter space explored in this paper, the $a$ decay is always prompt.  We restrict our attention to $m_a > m_\tau + m_e$; as shown in Fig.~\ref{fig:Ctaue} below, other bounds on the $|C_{\tau e}|$ coupling become much stronger for $m_a < m_\tau$.

\section{Cross-section calculation}

\begin{figure}[t!]
     \centering
         \includegraphics[trim = {5cm 19cm 3.6cm 4cm}, clip, width=0.7\columnwidth]{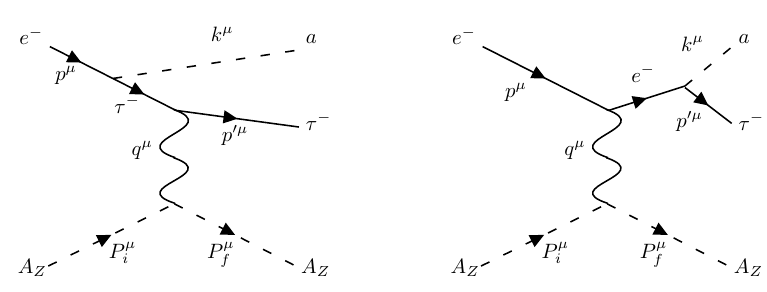}
     \caption{The diagrams which contribute to the process $eA_Z \rightarrow \tau A_Za$, where $A_Z$ is the ion nucleus.}
     \label{fig:diagram}
\end{figure}

We are now in a position to evaluate the diagram in \fig{fig:diagram}. The incoming momenta are denoted as $p^\mu$ for the electron and $P_i^\mu$ for the incoming ion, and the outgoing momenta are denoted as $p'^\mu$ for the $\tau$, $k$ for the ALP, and $P_f^\mu$ for the outgoing ion.  We restrict our attention to events in which the interaction with the ion is coherent and elastic, {\it i.e.} the ion does not break apart; these events will give the dominant contribution to our signal due to the $Z^2$ enhancement discussed below.  The four-momentum of the exchanged  photon is given by $q^\mu \equiv P_i^\mu - P_f^\mu$. For simplicity (following Refs. \cite{Liu:2016mqv,Liu:2017htz}), we will assume that the ion is a scalar boson with mass $M$, atomic number $Z$, and form factor $F(q^2)$, so that the interaction of a photon with the ion is
\bea
    iV^\mu(q^2, P_i, P_f) &= ieZF(q^2)(P_i^\mu + P_f^\mu).
\eea
From here forward, we will specialize to a gold ion ($Z=79$, mass number $A=197$), corresponding to a mass of $M=183$ GeV.  The form factor is an approximation of the Fourier transform of the Woods-Saxon distribution applied to the gold nucleus \cite{Klein:1999qj}, given by
\beq \label{eq:form_factor}
    F(q^2) = \frac{3}{q^3R_A^3}\left(\sin{qR_A} - qR_A\cos{qR_A}\right)\frac{1}{1 + a_0^2q^2}
\eeq
where $a_0 = 0.79~{\rm fm}$, $R_A = (1.1~{\rm fm})A^{1/3}$. For low momentum  transfer, $F(q^2) \approx 1$, so the amplitude is proportional to $Z$. As a result, the cross section will be enhanced by a factor of $79^2\approx 6000$ for small momentum transfer. This $Z^2$ enhancement enables the constraints made on $C_{\tau e}$ to be competitive with existing constraints, as we discuss below.  We have found, through numerical inspection, that $F(q)\sim 1$ until $m_a\sim 20$~GeV, corresponding to the region with the above $Z^2$ enhancement.  Above this mass scale, $F(q)<1$ and decreases with larger $m_a$, leading to a suppressed ALP production cross section. To restrict to the signal region where nuclear breakup does not occur, we additionally require a cutoff in the form factor of $q^2 < (100~{\rm MeV})^2$, corresponding to $m_a \sim 27~{\rm GeV}$.

To compute the amplitude, it is useful to define some Mandelstam-like variables in terms of the momenta of the diagram. We have
\begin{align}
    \tilde{s} &= (p'+k)^2 - m_e^2\\
    \tilde{u} &= (p-k)^2 - m_\tau^2\\
    t &= -q^2
\end{align}
The amplitude calculation is done in full detail in Appendix A; the final spin-averaged result is given by
\bea
    \overline{|\mathcal{M}|^2} = \left(\frac{4\pi Z \alpha C_{\tau e}m_\tau}{\Lambda}\right)^2\frac{F(q^2)^2}{q^4}\overline{|{\cal A}|^2}\,,
\eea
where $\alpha$ is the fine-structure constant, and
\begin{align}
\overline{|{\cal A}|^2} &= \frac{(\tilde{s}+\tilde{u})^2}{\tilde{s}\tilde{u}}P^2 - \frac{4t}{\tilde{s}\tilde{u}}(P\cdot k)^2 + \frac{(\tilde{s}+\tilde{u})^2}{\tilde{s}^2\tilde{u}^2}M^2(\theta_{\tau e})\left[P^2t - 4\left(\frac{\tilde{u}P\cdot p + \tilde{s}P\cdot p'}{\tilde{s}+\tilde{u}}\right)^2\right]
\end{align}
where $M^2(\theta_{\tau e}) = m_a^2 - m_\tau^2 - m_e^2 + 2m_\tau m_e \cos{(2\theta_{\tau e})}$. Note that for the spin-averaged amplitude, the only dependence on the parity violating angle $\theta_{\tau e}$ is an ${\cal O}(m_e/m_\tau)$ correction to the amplitude, so to very good precision one can compute the spin-averaged cross-section with $\overline{|{\cal A}_0|^2} \equiv \overline{|{\cal A}|^2}(\theta_{\tau e} = 0)$. We find that these results are in agreement with Refs.~\cite{Liu:2016mqv,Liu:2017htz} with the replacement $m_\tau \rightarrow m_e$ (any apparent sign discrepancies are due to the choice of the metric, which we take to be mostly negative).

\begin{figure}[t]
     \centering
         \includegraphics[ width=0.6\columnwidth]{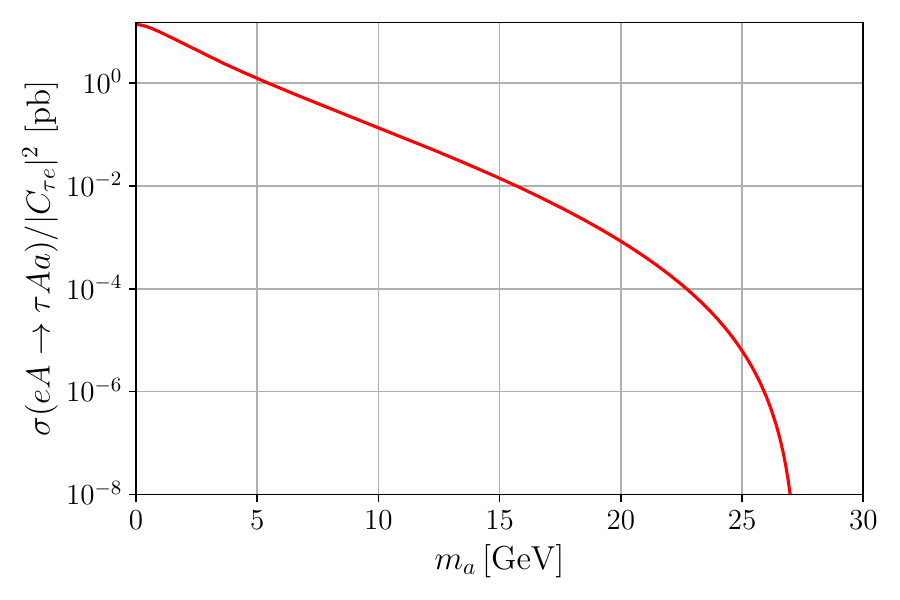}
     \caption{The cross section for the process $eA_Z \rightarrow \tau A_Za$ as a function of $m_a$, assuming an ALP UV scale $\Lambda = 1 \tev$ and assuming a Woods-Saxon nuclear form factor, given in Eq.~\ref{eq:form_factor}.}
     \label{fig:crossx}
\end{figure}

To compute the cross section, the integral over phase space is also done following Ref.~\cite{Liu:2017htz} closely.  Defining $x = E_k/|\bf p|$ to be the fraction of energy the ALP has w.r.t. the electron (assuming $p^0 = |{\bf p}|$), the differential cross-section in the rest frame of the ion is given by
\beq
    \frac{d^2\sigma}{dxd\cos{\theta_k}} = \frac{|{\bf k}|}{(32\pi^2)^2 M^2V} \int_{t_-}^{t_+}
    \!\!\!\!dt
    \int_0^{2\pi}d\phi_q \overline{|\mathcal{M}|^2},\label{eq:diffcrossx}
\eeq
where $V = |{\bf V}| = |{\bf p} - {\bf k}|$, $\theta_k$ is the angle ${\bf k}$ makes with ${\bf p}$,  $\phi_q$ is the azimuthal angle of ${\bf q}$ about ${\bf V}$, and $t_{\pm}$ are kinematic bounds from the energy-conserving $\delta$-function in the Lorentz-invariant phase space.  Details on the phase-space calculation are given in the Appendix B.

In order to evaluate the phase-space integral, we must determine the values of the initial-state momenta in \fig{fig:diagram}. For this, we use Table 10.2 of the EIC Yellow Report \cite{AbdulKhalek:2021gbh}, which states that the highest energies for the electron and gold ion beams are $|{\bf p}^{\rm lab}| = 18\gev$ and $E_{i}^{\rm lab} = 110\gev/{\rm nucleon}$, respectively. In the rest frame of the gold ion, a relativistic calculation reveals that the electron momentum has a magnitude of $|{\bf p}| \approx 4200\gev$. 

Next, we must determine the range of integration for $\cos \theta_k$.  Following the detector requirements listed in Table 10.6 of the EIC Yellow Report, we assume a detector pseudorapidity range of $|\eta| < 3.5$, corresponding to a range of angles $0.04 < \theta^{\rm lab} < \pi - 0.04$ in the lab frame. For most of the phase-space of interest, the differential cross-section in Eq.~\ref{eq:diffcrossx} is highly peaked near $\theta = 0$, making numerical integration challenging. This can be remedied by transforming from $\cos{\theta_k}$ to $\eta$, with the added benefit that one can then directly integrate over the allowed pseudorapidity region. The details of this transformation are shown in the Appendix B.

After inserting the kinematic parameters and doing the relevant transformations described above, the integral can be evaluated. The $\phi_q$ integrals are computed analytically, and the rest are done via trapezoid integration. The result for the total cross-section integration are shown in \fig{fig:crossx}.

Finally, we comment on the effect of beam polarization on the ALP signal.  To compute the polarized amplitude or cross-section, there are slight modifications to be made. In particular, one can rewrite the $\tau e$ coupling as
\begin{align}
    {\cal L}_{\tau e} &\approx \frac{C_{\tau e}m_\tau}{\Lambda}a [ R(\theta_{\tau e})\tau_L^\dagger e_R + L(\theta_{\tau e})\tau_R^\dagger e_L] + h.c. \label{eq:lag_pol}
\end{align}
where $R(\theta) = \cos{\theta}-\sin{\theta}$ and $L(\theta) = \cos{\theta}+\sin{\theta}$. The left-polarized cross-section can then be obtained by singling out the left-handed piece of \eq{eq:lag_pol}. This simply corresponds to setting $\theta_{\tau e} = \pi/4$, then multiplying by $\frac{1}{\sqrt{2}}L(\theta_{\tau e})$. Hence, we find
\begin{align}
|{\cal A}_L|^2 &= \frac{1}{2}|L(\theta_{\tau e})|^2\overline{|{\cal A}(\theta_{\tau e} = \pi/4)|^2} \nonumber\\&\approx |L(\theta_{\tau e})|^2\overline{|{\cal A}_0|^2}\,,
\end{align}
where have used $\overline{|{\cal A}(\theta_{\tau e} = \pi/4)|^2} = 2\overline{|{\cal A}(\theta_{\tau e} = \pi/4)|^2} \approx 2\overline{|{\cal A}_0|^2}$. The additional factor of 2 accounts for the fact that $\overline{|{\cal A}|^2}$ is a spin-averaged quantity. We similarly obtain
\begin{align}
|{\cal A}_R|^2 &= \frac{1}{2}|R(\theta_{\tau e})|^2\overline{|{\cal A}(\theta_{\tau e} = 3\pi/4)|^2} \nonumber\\ &\approx |R(\theta_{\tau e})|^2\overline{|{\cal A}_0|^2}.
\end{align}
Hence, to good approximation, $\sigma_L = |L(\theta_{\tau e})|^2\sigma_0$ and $\sigma_R = |R(\theta_{\tau e})|^2\sigma_0$, where $\sigma_0$ represents the spin-averaged cross-section when $\theta_{\tau e} = 0$. This allows us to compute the left-right asymmetry: 
\bea
r_{LR}(\theta_{\tau e}) = \frac{\sigma_L - \sigma_R}{\sigma_L + \sigma_R} = \sin{2\theta_{\tau e}}
\eea
Observation of an ALP signal at varying beam polarizations can thus be used as a direct probe of the parity-violating angle $\theta_{\tau e}$.

\section{Results\label{sec:results}}

Once the cross section is determined for a large sample of masses, limits can be placed on the coupling $C_{\tau e}$. Since the process always has an $a e\tau$ vertex, the cross section is proportional to $|C_{\tau e}|^2/\Lambda^2$. As a result, one can write the cross section as
\beq
\sigma(eA_Z\rightarrow \tau A_Z a) \equiv \frac{|C_{\tau e}|^2}{\Lambda^2}\hat{\sigma}.
\eeq
In what follows, we will assume that the ALP always decays leptonically in the detector and that $m_a > m_\tau$. We note that the addition of the coupling $C_{\tau \mu}$ may enhance the ability to detect LFV at the EIC, but we avoid this scenario for more straightforward comparison with the LFV constraints from Ref.~\cite{Cornella:2019uxs} (and the possible lack of a dedicated muon detection capability, as mentioned earlier). Due to the mass-dependence of the ALP coupling, there are only three significant decay channels for the ALP: (i) $a\rightarrow \tau^-\tau^+$, (ii) $a \rightarrow \tau^- e^+$ and (iii) $a\rightarrow \tau^+e^-$. 

Here, we note that the most advantageous aspect of our proposal for investigating LFV at the EIC is the ability to probe the $C_{\tau e}$ coupling nearly independently of the flavor-diagonal $C_{\tau\tau}$ coupling. In particular, when the latter is relatively suppressed, we find that the EIC can provide a promising venue for accessing the former. In that case, the final states of interest are $a(\rightarrow e^{\pm}\tau^{\mp})\tau^-$, on which we will focus. In particular, for our search, we will require the identification of an $e^+$ and a $\tau^{\pm}$ in the final state, and we will veto on the identification of an $e^-$. The largest source of irreducible background would then arise from electromagnetic production of $\tau$ pairs, which will be dominated by production through the Bethe-Heitler process \cite{Bethe:1934za,Ganapathi:1978qm,Akhundov:1979bd}.  To estimate this cross-section, we rescale the cross-sections found in Ref.~\cite{Bulmahn:2008fa}, which investigates the cross-section of ditau production from high-energy muons. We expect the difference between muon and electron collisions to be negligible for sufficiently high incident energies.  We take the results are found for ``rock'' ($Z=11$, $A=22$) and rescale the cross-section by $(Z_{\rm Au}/Z_{\rm rock})^2 = 51$. For an incident beam energy of $E = 4200\,{\rm GeV}$, this corresponds to a cross-section of $\sigma_{\rm b.g.} = 2.6\times 10^4~{\rm pb}$. For $\tau$ identification, we will adopt the $\tau$ efficiency found in Ref.~\cite{Zhang:2022zuz} of $\epsilon_\tau\approx 1\%$, which only considers 3-pronged decays, though we note that this is completely ignoring the other $\tau$ decay modes and an improved analysis could likely give a higher efficiency.  We also expect that one could improve the efficiency from the 3-pronged channel by vetoing on breakup of the ion, since this choice has no effect on our signal but reduces hadronic background. To calculate the background efficiency, we take a rate of $\sim 10^{-2}$ for losing the initial-state electron completely \cite{AbdulKhalek:2021gbh} and a rate of $\sim 10^{-3}$ for misidentification of the $e^-$ as an $e^+$, in line with studies of pions faking an electron from \cite{AbdulKhalek:2021gbh}.  Then, the overall background efficiency is
\begin{equation}
    \epsilon_{\rm b.g.} = 10^{-3}\cdot 10^{-2}\cdot (1 - 0.18) + 10^{-2}\cdot 10^{-2}\cdot 0.18 = 3.6\times 10^{-5},
\end{equation}
where the first term comes from misidentifying the electron as a positron (and assumes the $\tau^-$ does not decay to $e^-$), and the second term comes from losing the electron down the beam-pipe and detecting a positron from the decay of the $\tau^+$. Hence, the number of expected background events with ${\cal L} = (100/A)~{\rm fb}^{-1}$ of integrated luminosity is $n_{\rm b.g.} = \epsilon_{\rm b.g.}\sigma_{\rm b.g.}{\cal L} \approx 475$. Following the analysis done in Ref.~\cite{Feldman:1997qc}, the upper end of the 90\% confidence interval for Poisson signal mean given 475 mean background events and 475 observed events is $n_{\rm max} = 35$. Hence, for a signal with acceptance$\times$efficiency $\epsilon$, a value of $C_{\tau e}$ can be ruled out at a 90\% C.L. if
\begin{align}
\frac{|C_{\tau e}|^2}{\Lambda^2}\epsilon \sigma {\cal L} \geq  n_{\rm max}.
\end{align}
Here, the signal efficiency $\epsilon$ can be written as
\begin{align}
\epsilon = \epsilon_1{\cal B}(a \rightarrow e^+\tau^-) + \epsilon_2{\cal B}(a \rightarrow \tau^+\tau^-)
\end{align}
where $\epsilon_i$ represent the individual efficiencies of each possible branching. Adopting the same $\tau$ efficiency and positron misidentification rate, we have $\epsilon_1 = 2(0.01)(0.82) \approx 0.016$, and $\epsilon_2 = 2(0.01)(0.82)(0.18) \approx 0.003$.

\begin{figure*}[t!]
     \centering
         \includegraphics[width=\linewidth]{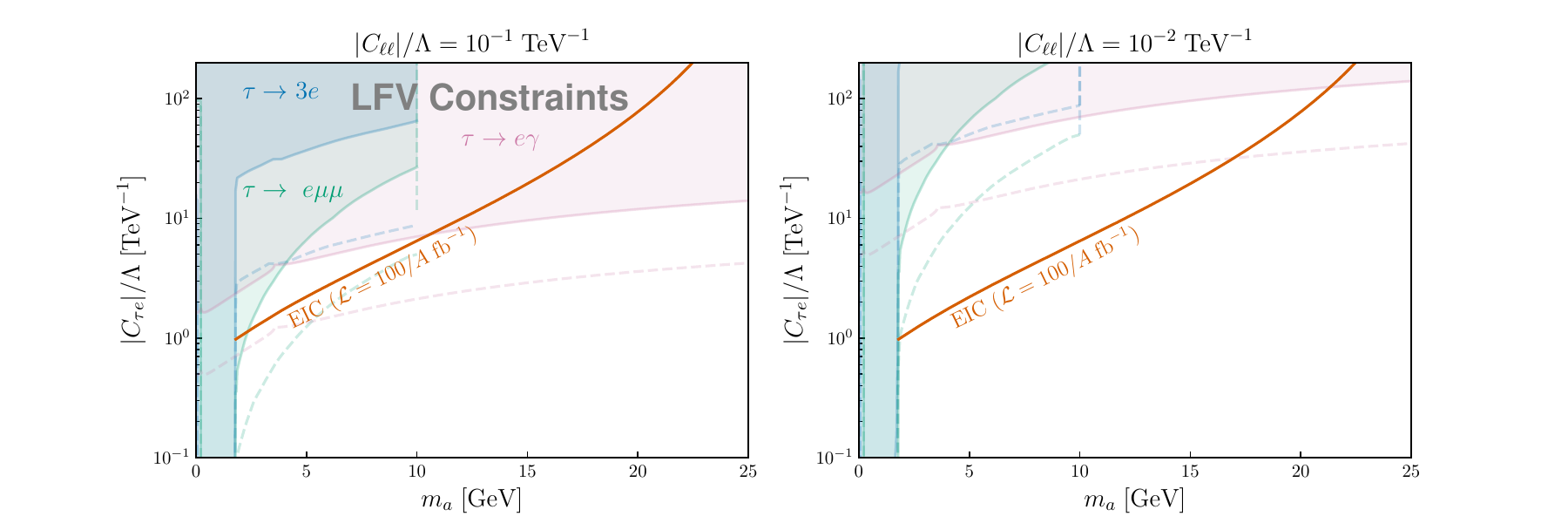}
     \caption{Projected constraints (90\% CL) on the interaction strength $C_{\tau e}$ from the EIC (with 100/A~fb$^{-1}$ of integrated ion luminosity) compared to LFV constraints from BABAR (solid lines) and projections from Belle II (with 50~ab$^{-1}$ of integrated luminosity, dashed lines) for $|C_{\ell\ell}|/\Lambda = 10^{-1} \tev^{-1}$ (left) and $|C_{\ell\ell}|/\Lambda = 10^{-2}\tev^{-1}$ (right), assuming that the only non-zero off-diagonal coupling is $C_{\tau e}$. The LFV limits (90\% CL) are taken from Ref.~\cite{Cornella:2019uxs} for $ m_a \leq 10\gev$, while the $\tau \rightarrow e\gamma$ limit is calculated explicitly for $m_a > 10\gev$ with formulae from Ref.~\cite{Cornella:2019uxs}, since this limit dominates in that regime.  The LFV limits are scaled up according to their dependence on $|C_{\ell \ell}| / \Lambda$, assuming that any contribution from the tree-level $a\gamma \gamma$ coupling is negligible.} 
     \label{fig:Ctaue}
\end{figure*}

The panels in \fig{fig:Ctaue} show the effect of assuming diagonal ALP couplings equal to $C_{\ell\ell} / \Lambda = 10^{-1}$ and $10^{-2}$ TeV${}^{-1}$.  Note that as $|C_{\ell \ell}|$ is reduced, our EIC production cross section remains  unchanged, since it is only  sensitive to $C_{\tau e}$.  As a result, our proposal for exploring the $C_{\tau e}$ LFV coupling can far exceed future projections using other probes, particularly when $C_{\ell \ell}$ is small and other indirect searches are less constraining.  The bounds obtained in the EIC search generally fall in a region of parameter space where $C_{\tau e}$ is significantly larger than $C_{\ell \ell}$; this may be realized in a general framework, or in the presence of significant parity violation as discussed above.

\section{Conclusions\label{sec:conclusions}}

Axion-like particles (ALPs) appear in a wide range of physical settings and may play a role in explaining some of the open questions of particle physics.  As such, they are well-motivated subjects of inquiry for theory and experiment.  In this work, we considered ALPs, at or above the GeV scale, whose couplings can lead to lepton-flavor-violation (LFV) and may even dominate their interactions with the SM.  We focused on $\tau-e$ LFV and showed that even with a fairly conservative analysis that does not make use of detailed kinematic information, the planned Electron-Ion Collider (EIC) can provide useful limits for this interaction, particularly in the case where the lepton-flavor-conserving couplings are suppressed.  This is mostly due to two factors: {\it (i)} a significant coherent enhancement of low-momentum-transfer electromagnetic scattering from a large $Z$ ion, mediating ALP emission, and {\it (ii)} the sizeable center-of-mass energy $\sim 100$~GeV envisioned for the EIC allowing it to reach for ALPs well above the GeV scale.  The final reach of such a search will depend on the EIC detectors' efficiency for $\tau$ identification, which may be significantly improved if muon detection capabilities are included in their final design \cite{Adkins:2022jfp}. 

The EIC can reach beyond current and projected bounds for $\tau-e$ LFV in tau decays for ALP masses above $m_\tau$ up to $m_a \sim 20$~GeV, assuming universal diagonal lepton couplings of order $|C_{\ell\ell}|/\Lambda \lsim 0.1\tev^{-1}$, with (100/A)~fb$^{-1}$ of gold ion scattering data.  The kinematics of the EIC are particularly favorable for production of ALPs towards the heavier end of this mass range, as opposed to fixed-target experiments \cite{Bjorken:2009mm, Liu:2017htz} which benefit from high luminosity and similar $Z^2$ enhancement, but have much lower collision energy.  A possible future fixed-target experiment using an ion beam at the LHC \cite{Hadjidakis:2018ifr} could possibly provide competitive limits to the EIC and is worth future study.

One notable observation is that the limits on $C_{\tau e}$ from the EIC are more robust and model-independent than other limits. Although we have focused on LFV constraints, this is generally true. For example, similar constraints were found for a universal $a\ell\ell'$ coupling by analyzing Higgs decays at the LHC in Ref.~\cite{Davoudiasl:2021haa}. These constraints were significantly weakened as the ALP-Higgs coupling was decreased, whereas the limits from the EIC would remain unaffected.

We also considered the possibility that ALPs may mediate parity-violating interactions.  This, for example, can be realized in certain models with composite ALPs from new strong dynamics.  Here, the EIC electron beam polarization can be a powerful probe of such interactions, making it a unique tool for illuminating the physics underlying LFV processes.

In principle, one could consider probing the muon-electron coupling $C_{\mu e}$ at the EIC using a similar search.  However, the characteristic mass dependence of ALP couplings reduces the corresponding cross section by $(m_\mu / m_\tau)^2$, so that an EIC search would not typically be competitive with other LFV  bounds \cite{Cornella:2019uxs} on $C_{\mu e}$, barring very small lepton-flavor diagonal $a\ell \ell$ and $a\gamma \gamma$ couplings.  Production of ALPs via the third LFV coupling $C_{\tau \mu}$ is not accessible in electron-ion collisions, but if both  $C_{\tau e}$ and $C_{\tau \mu}$ are significant then $a \rightarrow \mu \tau$ decays could provide access to the latter coupling, assuming that muon detection capabilities are included in the EIC detectors.

\section*{Acknowledgements}
We thank Nicholas Miesch for helpful discussions in the early stages of this work, and Craig Woody for useful comments.  This work is supported by the U.S. Department of Energy under Grant Contracts DE-SC0012704 (H.~D.) and DE-SC0010005 (E.~N. and R.~M.).

\bibliographystyle{JHEP}
\bibliography{eic-axions}

\providecommand{\href}[2]{#2}\begingroup\raggedright\begin{thebibliography}{10}

\bibitem{Peccei:1977hh}
R.~D. Peccei and H.~R. Quinn, {\it {CP Conservation in the Presence of
  Instantons}},  {\em Phys. Rev. Lett.} {\bf 38} (1977) 1440--1443.

\bibitem{Weinberg:1977ma}
S.~Weinberg, {\it {A New Light Boson?}},  {\em Phys. Rev. Lett.} {\bf 40}
  (1978) 223--226.

\bibitem{Wilczek:1977pj}
F.~Wilczek, {\it {Problem of Strong $P$ and $T$ Invariance in the Presence of
  Instantons}},  {\em Phys. Rev. Lett.} {\bf 40} (1978) 279--282.

\bibitem{Georgi:1986df}
H.~Georgi, D.~B. Kaplan, and L.~Randall, {\it {Manifesting the Invisible Axion
  at Low-energies}},  {\em Phys. Lett. B} {\bf 169} (1986) 73--78.

\bibitem{Wilczek:1982rv}
F.~Wilczek, {\it {Axions and Family Symmetry Breaking}},  {\em Phys. Rev.
  Lett.} {\bf 49} (1982) 1549--1552.

\bibitem{Ema:2016ops}
Y.~Ema, K.~Hamaguchi, T.~Moroi, and K.~Nakayama, {\it {Flaxion: a minimal
  extension to solve puzzles in the standard model}},  {\em JHEP} {\bf 01}
  (2017) 096, [\href{http://arxiv.org/abs/1612.05492}{{\tt arXiv:1612.05492}}].

\bibitem{Bauer:2019gfk}
M.~Bauer, M.~Neubert, S.~Renner, M.~Schnubel, and A.~Thamm, {\it {Axionlike
  Particles, Lepton-Flavor Violation, and a New Explanation of $a_\mu$ and
  $a_e$}},  {\em Phys. Rev. Lett.} {\bf 124} (2020), no.~21 211803,
  [\href{http://arxiv.org/abs/1908.00008}{{\tt arXiv:1908.00008}}].

\bibitem{Cornella:2019uxs}
C.~Cornella, P.~Paradisi, and O.~Sumensari, {\it {Hunting for ALPs with Lepton
  Flavor Violation}},  {\em JHEP} {\bf 01} (2020) 158,
  [\href{http://arxiv.org/abs/1911.06279}{{\tt arXiv:1911.06279}}].

\bibitem{Endo:2020mev}
M.~Endo, S.~Iguro, and T.~Kitahara, {\it {Probing $e\mu$ flavor-violating ALP
  at Belle II}},  {\em JHEP} {\bf 06} (2020) 040,
  [\href{http://arxiv.org/abs/2002.05948}{{\tt arXiv:2002.05948}}].

\bibitem{Iguro:2020rby}
S.~Iguro, Y.~Omura, and M.~Takeuchi, {\it {Probing $\mu\tau$ flavor-violating
  solutions for the muon $g-2$ anomaly at Belle II}},  {\em JHEP} {\bf 09}
  (2020) 144, [\href{http://arxiv.org/abs/2002.12728}{{\tt arXiv:2002.12728}}].

\bibitem{Han:2020dwo}
C.~Han, M.~L. L\'opez-Ib\'a\~nez, A.~Melis, O.~Vives, and J.~M. Yang, {\it
  {Anomaly-free leptophilic axionlike particle and its flavor violating
  tests}},  {\em Phys. Rev. D} {\bf 103} (2021), no.~3 035028,
  [\href{http://arxiv.org/abs/2007.08834}{{\tt arXiv:2007.08834}}].

\bibitem{Mukaida:2021sgv}
K.~Mukaida, K.~Schmitz, and M.~Yamada, {\it {Leptoflavorgenesis: baryon
  asymmetry of the Universe from lepton flavor violation}},
  \href{http://arxiv.org/abs/2111.03082}{{\tt arXiv:2111.03082}}.

\bibitem{Bauer:2021mvw}
M.~Bauer, M.~Neubert, S.~Renner, M.~Schnubel, and A.~Thamm, {\it {Flavor probes
  of axion-like particles}},  \href{http://arxiv.org/abs/2110.10698}{{\tt
  arXiv:2110.10698}}.

\bibitem{Davoudiasl:2017zws}
H.~Davoudiasl, P.~P. Giardino, E.~T. Neil, and E.~Rinaldi, {\it {Unified
  Scenario for Composite Right-Handed Neutrinos and Dark Matter}},  {\em Phys.
  Rev. D} {\bf 96} (2017), no.~11 115003,
  [\href{http://arxiv.org/abs/1709.01082}{{\tt arXiv:1709.01082}}].

\bibitem{Davoudiasl:2021haa}
H.~Davoudiasl, R.~Marcarelli, N.~Miesch, and E.~T. Neil, {\it {Searching for
  Flavor-Violating ALPs in Higgs Decays}},
  \href{http://arxiv.org/abs/2105.05866}{{\tt arXiv:2105.05866}}.

\bibitem{ParticleDataGroup:2020ssz}
{\bf Particle Data Group} Collaboration, P.~A. Zyla et~al., {\it {Review of
  Particle Physics}},  {\em PTEP} {\bf 2020} (2020), no.~8 083C01.

\bibitem{Gonderinger:2010yn}
M.~Gonderinger and M.~J. Ramsey-Musolf, {\it {Electron-to-Tau Lepton Flavor
  Violation at the Electron-Ion Collider}},  {\em JHEP} {\bf 11} (2010) 045,
  [\href{http://arxiv.org/abs/1006.5063}{{\tt arXiv:1006.5063}}]. [Erratum:
  JHEP 05, 047 (2012)].

\bibitem{Cirigliano:2021img}
V.~Cirigliano, K.~Fuyuto, C.~Lee, E.~Mereghetti, and B.~Yan, {\it {Charged
  Lepton Flavor Violation at the EIC}},  {\em JHEP} {\bf 03} (2021) 256,
  [\href{http://arxiv.org/abs/2102.06176}{{\tt arXiv:2102.06176}}].

\bibitem{Zhang:2022zuz}
J.~L. Zhang et~al., {\it {Search for $e\to\tau$ Charged Lepton Flavor Violation
  at the EIC with the ECCE Detector}},
  \href{http://arxiv.org/abs/2207.10261}{{\tt arXiv:2207.10261}}.

\bibitem{Furletova:2021wyq}
Y.~Furletova and S.~Mantry, {\it {Probing charged lepton flavor violation with
  a positron beam at CEBAF (JLAB)}},
  \href{http://arxiv.org/abs/2111.03912}{{\tt arXiv:2111.03912}}.

\bibitem{Liu:2021lan}
Y.~Liu and B.~Yan, {\it {Searching for the axion-like particle at the EIC}},
  \href{http://arxiv.org/abs/2112.02477}{{\tt arXiv:2112.02477}}.

\bibitem{Bauer:2017ris}
M.~Bauer, M.~Neubert, and A.~Thamm, {\it {Collider Probes of Axion-Like
  Particles}},  {\em JHEP} {\bf 12} (2017) 044,
  [\href{http://arxiv.org/abs/1708.00443}{{\tt arXiv:1708.00443}}].

\bibitem{Bauer:2020jbp}
M.~Bauer, M.~Neubert, S.~Renner, M.~Schnubel, and A.~Thamm, {\it {The
  Low-Energy Effective Theory of Axions and ALPs}},
  \href{http://arxiv.org/abs/2012.12272}{{\tt arXiv:2012.12272}}.

\bibitem{Chala:2020wvs}
M.~Chala, G.~Guedes, M.~Ramos, and J.~Santiago, {\it {Running in the ALPs}},
  {\em Eur. Phys. J. C} {\bf 81} (2021), no.~2 181,
  [\href{http://arxiv.org/abs/2012.09017}{{\tt arXiv:2012.09017}}].

\bibitem{Escribano:2020wua}
P.~Escribano and A.~Vicente, {\it {Ultralight scalars in leptonic
  observables}},  {\em JHEP} {\bf 03} (2021) 240,
  [\href{http://arxiv.org/abs/2008.01099}{{\tt arXiv:2008.01099}}].

\bibitem{Calibbi:2020jvd}
L.~Calibbi, D.~Redigolo, R.~Ziegler, and J.~Zupan, {\it {Looking forward to
  Lepton-flavor-violating ALPs}},  \href{http://arxiv.org/abs/2006.04795}{{\tt
  arXiv:2006.04795}}.

\bibitem{Ma:2021jkp}
K.~Ma, {\it {Polarization Effects in Lepton Flavor Violated Decays Induced by
  Axion-Like Particles}},  \href{http://arxiv.org/abs/2104.11162}{{\tt
  arXiv:2104.11162}}.

\bibitem{Liu:2016mqv}
Y.-S. Liu, D.~McKeen, and G.~A. Miller, {\it {Validity of the
  Weizs\"acker-Williams approximation and the analysis of beam dump
  experiments: Production of a new scalar boson}},  {\em Phys. Rev. D} {\bf 95}
  (2017), no.~3 036010, [\href{http://arxiv.org/abs/1609.06781}{{\tt
  arXiv:1609.06781}}].

\bibitem{Liu:2017htz}
Y.-S. Liu and G.~A. Miller, {\it {Validity of the Weizs\"acker-Williams
  approximation and the analysis of beam dump experiments: Production of an
  axion, a dark photon, or a new axial-vector boson}},  {\em Phys. Rev. D} {\bf
  96} (2017), no.~1 016004, [\href{http://arxiv.org/abs/1705.01633}{{\tt
  arXiv:1705.01633}}].

\bibitem{Klein:1999qj}
S.~Klein and J.~Nystrand, {\it {Exclusive vector meson production in
  relativistic heavy ion collisions}},  {\em Phys. Rev. C} {\bf 60} (1999)
  014903, [\href{http://arxiv.org/abs/hep-ph/9902259}{{\tt hep-ph/9902259}}].

\bibitem{AbdulKhalek:2021gbh}
R.~Abdul~Khalek et~al., {\it {Science Requirements and Detector Concepts for
  the Electron-Ion Collider: EIC Yellow Report}},
  \href{http://arxiv.org/abs/2103.05419}{{\tt arXiv:2103.05419}}.

\bibitem{Bethe:1934za}
H.~Bethe and W.~Heitler, {\it {On the Stopping of fast particles and on the
  creation of positive electrons}},  {\em Proc. Roy. Soc. Lond. A} {\bf 146}
  (1934) 83--112.

\bibitem{Ganapathi:1978qm}
V.~Ganapathi and J.~Smith, {\it {Electromagnetic Production of Trimuons in Deep
  Inelastic Muon Scattering}},  {\em Phys. Rev. D} {\bf 19} (1979) 801--809.

\bibitem{Akhundov:1979bd}
A.~A. Akhundov, D.~Y. Bardin, N.~D. Gagunashvili, and N.~M. Shumeiko, {\it
  {Electromagnetic Trident Production in Deep Inelastic $\mu N$ Scattering}},
  {\em Sov. J. Nucl. Phys.} {\bf 31} (1980) 127.

\bibitem{Bulmahn:2008fa}
A.~Bulmahn and M.~H. Reno, {\it {Cross sections and energy loss for lepton pair
  production in muon transport}},  {\em Phys. Rev. D} {\bf 79} (2009) 053008,
  [\href{http://arxiv.org/abs/0812.5008}{{\tt arXiv:0812.5008}}].

\bibitem{Feldman:1997qc}
G.~J. Feldman and R.~D. Cousins, {\it {A Unified approach to the classical
  statistical analysis of small signals}},  {\em Phys. Rev. D} {\bf 57} (1998)
  3873--3889, [\href{http://arxiv.org/abs/physics/9711021}{{\tt
  physics/9711021}}].

\bibitem{Adkins:2022jfp}
J.~K. Adkins et~al., {\it {Design of the ECCE Detector for the Electron Ion
  Collider}},  \href{http://arxiv.org/abs/2209.02580}{{\tt arXiv:2209.02580}}.

\bibitem{Bjorken:2009mm}
J.~D. Bjorken, R.~Essig, P.~Schuster, and N.~Toro, {\it {New Fixed-Target
  Experiments to Search for Dark Gauge Forces}},  {\em Phys. Rev. D} {\bf 80}
  (2009) 075018, [\href{http://arxiv.org/abs/0906.0580}{{\tt
  arXiv:0906.0580}}].

\bibitem{Hadjidakis:2018ifr}
C.~Hadjidakis et~al., {\it {A fixed-target programme at the LHC: Physics case
  and projected performances for heavy-ion, hadron, spin and astroparticle
  studies}},  {\em Phys. Rept.} {\bf 911} (2021) 1--83,
  [\href{http://arxiv.org/abs/1807.00603}{{\tt arXiv:1807.00603}}].

\end{thebibliography}\endgroup


\appendix

\section{Amplitude Calculation}

In computing the amplitude, we borrow notation from Refs. \cite{Liu:2016mqv,Liu:2017htz}. In this appendix, we attempt to provide the amplitude calculation in as much detail as possible, stating explicitly whenever a computer algebra system was used.

Let the incoming four-momenta be $p$ for the electron and $P_i$ for the incoming ion. Let the outgoing momenta be $p'$ for the $\tau$, $k$ for the ALP, and $P_f$ for the outgoing ion. Also, let the mass of the ALP be $m_a$, the mass of the ion be $M$, and the charge of the ion be $Z$. We define $P \equiv P_i + P_f$ and $q \equiv P_i - P_f$, along with the following Mandelstam variables:
\begin{align}
    \tilde{s} &= (p'+k)^2 - m_e^2 = 2p'\cdot k + m_a^2 + m_\tau^2 - m_e^2\label{eq:mandelstam1}\\
    \tilde{u} &= (p-k)^2 - m_\tau^2 =-2p\cdot k + m_a^2 + m_e^2 - m_\tau^2\label{eq:mandelstam2}\\
    t_2 &= (p'-p)^2 = -2p'\cdot p + m_e^2 + m_\tau^2\label{eq:mandelstam3}\\
    t &= -q^2,\label{eq:mandelstam4}
\end{align}
which satisfy $\tilde{s}+t_2 + \tilde{u}+t = m_a^2$.

The diagrams in Fig.\ref{fig:diagram} are relatively straightforward, and yield
\begin{align}
    i\mathcal{M}_1 &= \bar{u}(p')ie\gamma_\mu\frac{i}{\slashed{p}-\slashed{k} - m_\tau}i(\sin{\theta_{\tau e}} - \gamma_5\cos{\theta_{\tau e}})\frac{m_\tau C_{\tau e}}{\Lambda}u(p)\frac{V^\mu(q^2, P_i, P_f)}{q^2},\\
    i\mathcal{M}_2 &= \bar{u}(p')i(\sin{\theta_{\tau e}} - \gamma_5\cos{\theta_{\tau e}})\frac{m_\tau C_{\tau e}}{\Lambda}\frac{i}{\slashed{p}'+\slashed{k} - m_e}ie\gamma_\mu u(p)\frac{V^\mu(q^2, P_i, P_f)}{q^2}\,,
\end{align}
where 
\begin{equation}
    iV^\mu(q^2, P_i, P_f) = ieZF(q^2)(P_i^\mu + P_f^\mu)
\end{equation}
is the photon-ion interaction vertex. The total amplitude can then be written as
\begin{align}
    i{\cal M} &= \frac{4\pi Z\alpha C_{\tau e}m_\tau}{\Lambda}\frac{F(q^2)}{q^2}P^\mu \bar{u}(p')\Gamma_\mu(p, k, k')u(p)\,,
\intertext{where}
    \Gamma_\mu(p, k, k') &= \left[\gamma^\mu\frac{\slashed{p} - \slashed{k} + m_\tau}{(p - k)^2-m_\tau^2} + \frac{\slashed{p}'+\slashed{k} - m_e}{(p' + k)^2 - m_e^2}\gamma^\mu\right]\sin{\theta_{\tau e}}\\
    &-\left[\gamma^\mu\frac{\slashed{p} - \slashed{k} + m_\tau}{(p - k)^2-m_\tau^2} - \frac{\slashed{p}'+\slashed{k} + m_e}{(p' + k)^2 - m_e^2}\gamma^\mu\right]\gamma_5\cos{\theta_{\tau e}}.
\end{align}
The spin-averaged squared amplitude is then given by
\begin{align}
    \overline{|\mathcal{M}|^2} &= \left(\frac{4\pi Z \alpha C_{\tau e}m_\tau}{\Lambda}\right)^2\frac{F(q^2)^2}{q^4}\overline{|{\cal A}|^2}\,,
\end{align}
with 
\begin{align}
    \overline{|{\cal A}|^2} &=P^\mu P^\nu\frac{1}{2}\sum_{\sigma \sigma'}\bar{u}_{\sigma}(p)\Gamma_{\mu}^{\dagger}(p, k, k')u_{\sigma'}(p')\bar{u}_{\sigma'}(p')\Gamma_\nu(p,k,k')u_{\sigma}(p)\\
    &= \frac{1}{2}P^\mu P^\nu {\rm tr}\left\{(\slashed{p} + m_e)\Gamma_\mu^\dagger(p, k, k') (\slashed{p}' + m_\tau)\Gamma_\nu(p, k, k')\right\}.
\end{align}
One can compute this trace in a computer algebra system, then simplify using the Mandelstam variables defined in Eqs.~(\ref{eq:mandelstam1}-\ref{eq:mandelstam4}). It is given by
\begin{equation}
\overline{|{\cal A}|^2} = \frac{(\tilde{s}+\tilde{u})^2}{\tilde{s}\tilde{u}}P^2 - 4\frac{t}{\tilde{s}\tilde{u}}(P\cdot k)^2 + \frac{(\tilde{s}+\tilde{u})^2}{\tilde{s}^2\tilde{u}^2}M^2(\theta)\left[P^2t - 4\left(\frac{\tilde{u}P\cdot p + \tilde{s}P\cdot p'}{\tilde{s}+\tilde{u}}\right)^2\right]\,,
\label{eq:amplitude}
\end{equation}
where $M^2(\theta) = m_a^2 - m_\tau^2 - m_e^2 + 2m_\tau m_e \cos{(2\theta)}$. Note that $M^2(\theta) \approx m_a^2 - m_\tau^2$, regardless of $\theta$.
\section{Differential Cross Section Integration}
In the initial-state ion rest frame, the differential cross-section is given by
\begin{align}
    d\sigma &= \frac{1}{4|{\bf{p}}|M}\overline{|\mathcal{M}|^2}(2\pi)^4\delta^4(p'+k-p-q)\frac{d^3{{\bf p}'}}{(2\pi)^32E'}\frac{d^3{\bf{P}}_f}{(2\pi)^32E_f}\frac{d^3{\bf{k}}}{(2\pi)^32E_k}.
\end{align}
 where $E'$, $E_f$, and $E_k$ are the time-components of the four-momenta $p'$, $P_f$, and $k$, respectively. The first step in simplifying $d\sigma$ is converting variables from ${\bf P}_f$ to ${\bf q}$ (which has unit Jacobian) and integrating over ${\bf p}'$. Doing so yields
\begin{align}
    d\sigma &= \frac{\overline{|\mathcal{M}|^2}}{1024\pi^5|{\bf p}|ME_f E'E_k}\delta(E' + E_k - E - q_0)d^3{\bf q}d^3{\bf k}
\end{align}
In order to simplify some expressions, let ${\bf V} = {\bf p} - {\bf k}$ and $V = |{\bf V}|$, and define ${\bf q} \equiv (Q, \theta_q, \phi_q)$ in the direction of ${\bf V}$. Then we find
\begin{align}
    E' &= \sqrt{Q^2 + V^2 + 2QV\cos{\theta_q} + m_\tau^2}\\
    E_f &= \sqrt{Q^2+M^2}\\
    E_k &= \sqrt{|{\bf k}|^2 + m_a^2}\\
    E &= \sqrt{|{\bf p}|^2 + m_e^2}\\
    q_0 &= M - \sqrt{Q^2 + M^2}.
\end{align}
One can replace $E_k$ systematically with $\tilde{u}$, since \begin{align}
    \tilde{u} &= (p-k)^2 - m_\tau^2\nonumber\\ 
    &= (E - E_k)^2 - |{\bf p}-{\bf k}|^2 - m_\tau^2\nonumber\\
    &= (E - E_k)^2 - V^2 - m_\tau^2\\
    \implies E_k &= E - \sqrt{\tilde{u} + V^2 + m_\tau^2}
\end{align}
Now we can integrate over $\theta_q$. To do this, note that the argument inside of the delta function
\begin{align}
    f(\cos{\theta_q}) &= E'(\cos{\theta_q}) + E_k - E - q_0
\end{align}
has derivative
\begin{align}
    f'(\cos{\theta_q}) &= \frac{QV}{\sqrt{Q^2+V^2 + 2QV\cos{\theta_q} + m_\tau^2}}
\end{align}
and a zero at
\begin{align}
    \cos{\theta_q^0} &= \frac{(E - E_k + q_0)^2 - m_\tau^2 - Q^2 - V^2}{2QV}.
\end{align}
Using this expression for $f'(\cos{\theta}_q)$ the delta function can be rewritten as
\begin{align}
    \frac{1}{E_f E' E_k}\,\delta(f(\cos{\theta_q})) &= \frac{1}{E_f E' E_k}\frac{1}{|f'(\cos{\theta_q^0})|}\,\delta(\cos{\theta_q} - \cos{\theta_q^0})\\
    &= \frac{1}{E_k Q\, V \sqrt{M^2+Q^2}}\,\delta(\cos{\theta_q} - \cos{\theta_q^0})
\end{align}
and the differential cross-section becomes
\begin{align}
    d\sigma &= \frac{\overline{|\mathcal{M}|^2}}{1024\pi^5|{\bf p}|M}\frac{1}{E_k Q\, V \sqrt{M^2+Q^2}}\,\delta(\cos{\theta_q} - \cos{\theta_q^0})Q^2 dQ d(\cos{\theta_q}) d\phi_q d^3{\bf k}.
\end{align}

Note that this solution is not always between $-1$ and $1$. For $|\cos{\theta}_q^0| > 1$, the process is kinematically forbidden, and this is enforced by the integral over the $\delta$-function. We can determine when this happens by solving for $\cos{\theta_q^0}(Q) = \pm 1$, which yields two positive and two negative solutions in $Q$. We only care about the positive solutions, which yields
\begin{align}
    Q_{\pm} &= \left|\frac{V[\tilde{u} + 2(E'+E_f)M] \pm (E' + E_f)\sqrt{\tilde{u}^2 + 4M(E' + E_f)\tilde{u} + 4M^2 V^2}}{2(E'+E_f)^2 - 2V^2}\right|.
\end{align}
As a result, we have
\begin{align}
    d\sigma &= \frac{d^3{\bf k}}{1024\pi^5|{\bf p}|VE_kM}\int_{Q_-}^{Q_+}dQ\frac{Q}{\sqrt{M^2+Q^2}}  d\phi_q \overline{|\mathcal{M}|^2}.
\end{align}
Alternatively, one can leave the integral over $Q$ unbounded, by noting that the integral of $\delta(\cos{\theta}_q - \cos{\theta}_q^0)$ with respect to $\cos{\theta}_q$ introduces a Heaviside $\Theta$ function, which automatically enforces the bounds:
\begin{align}
    d\sigma &= \frac{d^3{\bf k}}{1024\pi^5|{\bf p}|VE_kM}\int_{0}^{\infty}dQ\frac{Q}{\sqrt{M^2+Q^2}}  d\phi_q \overline{|\mathcal{M}|^2}\Theta(1 - \cos^2{\theta}_q^{0}).
\end{align}
This is the approach we take when evaluating the integral numerically. To simplify things further, we make a change of variables by introducing the Mandelstam variables $t$:
\begin{align}
    t = -q^2 &= Q^2 - (\sqrt{Q^2+M^2}-M)^2 \nonumber\\
            &= 2M(\sqrt{M^2+Q^2}-M).
\end{align}
This has $dt/dQ = 2MQ/\sqrt{M^2+Q^2}$, so
\begin{equation}
    d\sigma 
    = \frac{d^3{\bf k}}{128\pi^4|{\bf p}|VE_k}\int_{t(Q_-)}^{t(Q_+)}dt\left(\frac{1}{8M^2}\int_0^{2\pi}\frac{d\phi_q}{2\pi} \overline{|\mathcal{M}|^2}\right).
\end{equation}
We can now simplify the integral over $d^3{\bf k} = |{\bf k}|^2 d|{\bf k}|d\phi_k d(\cos{\theta}_k)$ by defining $x = E_k/E$, so that $dx/d|{\bf k}| = |{\bf k}|/EE_k$. Then,
\begin{align}
    \frac{d\sigma}{dx\,d(\cos{\theta_k})} 
    &= \frac{1}{64\pi^3}\frac{|{\bf k}|E}{|{\bf p}|V}\int_{t(Q_-)}^{t(Q_+)}dt\left(\frac{1}{8M^2}\int_0^{2\pi}\frac{d\phi_q}{2\pi} \overline{|\mathcal{M}|^2}\right)\\
    &\approx \frac{1}{64\pi^3}\frac{|{\bf k}|}{V}\int_{t(Q_-)}^{t(Q_+)}dt\left(\frac{1}{8M^2}\int_0^{2\pi}\frac{d\phi_q}{2\pi} \overline{|\mathcal{M}|^2}\right)
\end{align}
where we have taken $m_e \ll |{\bf p}|$.
It turns out that the integral over $\phi_q$ can be computed analytically. To do so, one must express the amplitude Eq.~(\ref{eq:amplitude}) in terms of the integration variables. We have
\begin{align}
    {\bf q}\cdot{\bf k}
    &=\frac{Q|{\bf k}|}{V}\left[|{\bf p}|(\cos{\theta_q^0}\cos{\theta_k} + \sin{\theta_q^0}\sin{\theta_k}\cos{\phi_q}) - |{\bf k}|\cos{\theta_q}\right]
\intertext{and}
    {\bf q}\cdot {\bf p}
    &= \frac{Q|{\bf p}|}{V}\left[|{\bf p}|\cos{\theta_q^0} - |{\bf k}|(\cos{\theta_q^0}\cos{\theta_k} - \sin{\theta_q^0}\sin{\theta_k}\cos{\phi_q})\right].
\end{align}
With these, the kinematic terms that appear in Eq.~(\ref{eq:amplitude}) can be represented in terms of the integration variables. We have
\begin{align}
    \tilde{s}&=  - \left(1 + \frac{E}{M}\right)t - 2({\bf q}\cdot{\bf p})\\
    \tilde{u} &= m_a^2 + m_e^2 - m_\tau^2 - 2xE^2 + 2|{\bf p}|\sqrt{x^2E^2 - m_a^2}\cos{\theta_k}\\
    P^2 & = 4M^2 + t\\
    P\cdot k &= \left(2 M + \frac{t}{2M}\right)xE + {\bf q}\cdot{\bf k}\\
    P\cdot p &= \left(2 M + \frac{t}{2M}\right)E + {\bf q}\cdot{\bf p}\\
    P\cdot p'
    &= P\cdot p - P\cdot k\,.
\end{align}
To compute the integral over $\phi_q$, we note that the only place $\phi_q$ appears is in the $\cos{\phi_q}$ inside of ${\bf q}\cdot{\bf k}$ and ${\bf q}\cdot{\bf p}$. As a result, each of the terms in Eq.~(\ref{eq:amplitude}) can be written in the form
\begin{equation}
    A + B\cos{\phi_q} + C\cos^2{\phi_q} + \frac{D}{F + G\cos{\phi_q}} + \frac{E}{(F + G\cos{\phi}_q)^2}\,.
\end{equation}
This can then be integrated according to
\begin{align}
    \int_0^{2\pi}{\frac{d\phi}{2\pi}\left[    A + B\cos{\phi} + C\cos^2{\phi} + \frac{D}{F + G\cos{\phi}} + \frac{E}{(F + G\cos{\phi})^2}\right]}\nonumber\\
    = A + \tfrac{1}{2}C + \frac{D}{(F^2 - G^2)^{1/2}} + \frac{EF}{(F^2-G^2)^{3/2}}
\end{align}
\begin{figure}[t]
     \centering
         \includegraphics[ width=0.7\columnwidth]{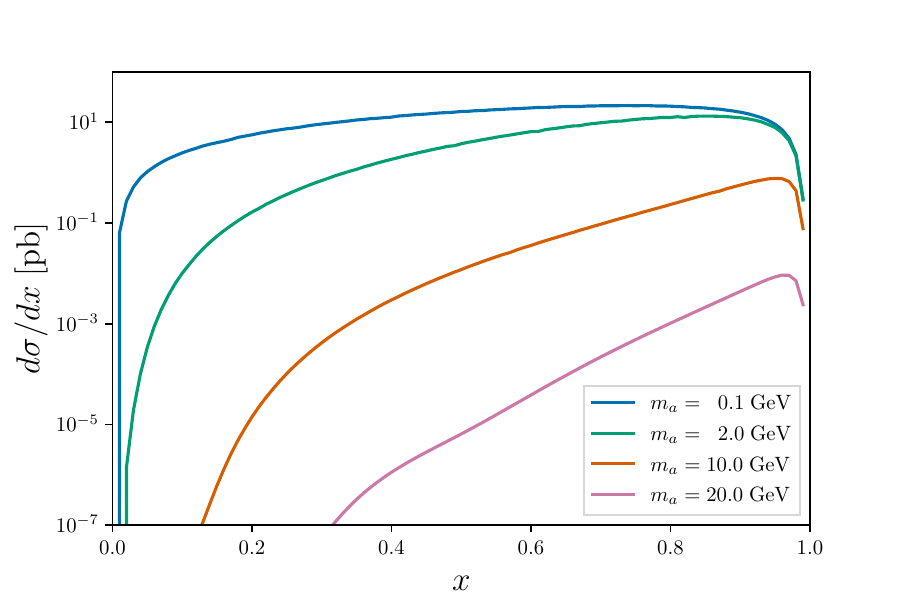}
     \caption{The differential cross section for the process $eA_Z \rightarrow \tau A_Z a$ as a function of the energy fraction $x = E_k/p$ for $m_a = 0.1,~2.0,~10.0, $ and $20.0 \gev$, assuming an interaction strength $|C_{\tau e}|/\Lambda = 1 \tev^{-1}$.  }
     \label{fig:dsig_dx}
\end{figure}
so in principle, $\int_0^{2\pi}\tfrac{d\phi_q}{2\pi}\overline{|{\cal A}|^2}$ can be computed analytically. The constants $A,B,C,D,$ and $E$ are complicated functions of the kinematic variables and differ term-by-term, so we do not write them out explicitly. However, this substitution can be made in a computer algebra system so that the only remaining integrals are over $t$, $\cos{\theta_k}$, and $x$. 

The integral over $\cos{\theta_k}$ is highly-peaked near $\cos{\theta_k} = 1$, but this can be simplified by converting to an integral over $\eta$ in the lab-frame. The angle of the ALP in the lab-frame is given by
\beq
    \tan{\theta_k^{\rm lab}} = \frac{\sin{\theta_k}}{\gamma_A(\cos{\theta_k} - v_A/u_k)}
\eeq
where $v_A$ is the speed of the ion in the lab-frame and $u_k = \sqrt{1 - m_a^2/E_k^2}$ is the speed of the ALP in the rest-frame of the ion. This can be solved for $\cos{\theta_k}$, yielding
\beq
\cos{\theta_k} = \frac{\sqrt{u_k^2+(u_k^2-v_A^2)\gamma_A^2\tan^2{\theta_k^{\rm lab}}}+v_A\gamma_A^2\tan^2{\theta_k^{\rm lab}}}{u_k(1+\gamma_A^2\tan^2{\theta_k^{\rm lab}})}\label{eq:costhk}
\eeq

Then, the pseudorapidity of the ALP in the lab-frame is given by
\beq
    \eta_k = -\log{(\tan(\theta_k^{\rm lab}/2)}).
\eeq
The differential cross-section is then given by
\begin{align}
    \frac{d\sigma}{dx\,d\eta_k} &= \frac{d\sigma}{dx\,d(\cos{\theta_k})}\frac{d(\cos{\theta_k})}{d\theta^{\rm lab}_k}\frac{d\theta^{\rm lab}_k}{d\eta_k}\\
    &= -\sin{\theta_k^{\rm lab}} \left[1-\frac{v_A}{\sqrt{u_k^2+(u_k^2-v_A^2)\gamma_A^2\tan^2{\theta_k^{\rm lab}}}}\right]\frac{\gamma_A^2\sec^2{\theta_k^{\rm lab}}}{1+\gamma_A^2\tan^2{\theta_k^{\rm lab}}}\frac{d\sigma}{dx\,d(\cos{\theta_k})}
\end{align}
where one must make the substitutions Eq.~\ref{eq:costhk} and $\theta_k^{\rm lab} = 2\arctan{e^{-\eta_k}}$ before performing the integral. This final result for $d\sigma/dx d\eta_k$ can then be integrated over $t$, $\eta_k$, and $x$ using the trapezoid rule. The results of integrating over $t$ and $\eta_k$ for a range of $m_a$ are shown in Fig.~\ref{fig:dsig_dx}.

\clearpage

\end{document}